# A biophysical protein folding model accounts for most mutational fitness effects in viruses

C. Scott Wylie and Eugene I. Shakhnovich
*Department of Chemistry and Chemical Biology, Harvard University.*
12 Oxford Street, Cambridge, MA 02138, USA

corresponding author: wylie@fas.harvard.edu

## *Abstract*

**Fitness effects of mutations fall on a continuum ranging from lethal to deleterious to beneficial. The distribution of fitness effects (DFE) among random mutations is an essential component of every evolutionary model and a mathematical portrait of robustness. Recent experiments on five viral species all revealed a characteristic bimodal shaped DFE, featuring peaks at neutrality and lethality. However, the phenotypic causes underlying observed fitness effects are still unknown, and presumably thought to vary unpredictably from one mutation to another. By combining population genetics simulations with a simple biophysical protein folding model, we show that protein thermodynamic stability accounts for a large fraction of observed mutational effects. We assume that moderately destabilizing mutations inflict a fitness penalty proportional to the reduction in folded protein, which depends continuously on folding free energy *($\Delta G$)*. Most mutations in our model affect fitness by altering *$\Delta G$*, while, based on simple estimates, $\approx$10% abolish activity and are unconditionally lethal. Mutations pushing *$\Delta G > 0$* are also considered lethal. Contrary to neutral network theory, we find that, in mutation/selection/drift steady-state, high mutation rates (*m*) lead to less stable proteins and a more dispersed DFE, i.e. less mutational robustness. Small population size (*N*) also decreases stability and robustness. In our model, a continuum of non-lethal mutations reduces fitness by $\approx$2% on average, while $\approx$10-35% of mutations are lethal, depending on *N* and *m*. Compensatory mutations are common in small populations with high mutation rates. More broadly, we conclude that interplay between biophysical and population genetic forces shapes the DFE.**

## *Introduction*

What fraction of new mutations is deleterious to organismal fitness? Are most deleterious mutations mild or are they nearly lethal? The answers to these fundamental questions are provided by the distribution of fitness effects (DFE). The DFE quantifies robustness of genomes to random mutations: deleterious mutations have small effects in robust genomes while having large or lethal effects in brittle genomes. The DFE also shapes the pattern and extent of genetic diversity segregating within populations. This diversity, in turn, is crucial to interpreting molecular polymorphism data (1), the evolutionary function of sex/recombination (2), and genomic decay due to "Muller's ratchet" (3). Finally, the DFE also constrains patterns of nucleotide substitutions between species, e.g. the "molecular clock" (4).

Properties of the DFE have long been estimated by two indirect methods. First, mutation accumulation experiments pass populations through deep bottlenecks, which relaxes selection and causes (mostly) deleterious mutations to accumulate, depressing the population's mean fitness (4). The rate and strength of typical mutations can be estimated from the tempo and variability of fitness decline. A second method compares the rate of nucleotide substitutions across species at sites of interest to that of putatively neutral sites (e.g. ref. (5)). Importantly, neither of these methods can detect lethal mutations because they are instantly purged from populations. For a review of these methods and the DFE generally, see ref. (6). Recently, a more direct method utilizing site-directed mutagenesis was applied to viruses (7-11). These studies measured mutant fitness paired with the exact underlying genomic change among an unbiased set of single nucleotide substitutions, finding similarly shaped DFE across five viral species.

Most missense mutations probably impact organismal fitness by altering protein activity and/or stability. Predicting which rare mutations dramatically improve protein activity or create new functions remains a formidable challenge that we do not address here. However, estimating the distribution of mutational effects that merely perturb evolved, more-or-less optimized proteins is a more tenable goal. The role of most residues is to maintain a protein's overall fold, and mutations at these sites mainly alter stability but not activity. The stability changes induced by those mutations are predictable in a statistical sense, as explained in Results. Though less predictable than stability, activity is governed mostly by just a few key residues, e.g. the active catalytic site, where mutations nearly always abolish activity. This general picture of protein organization underlies our biophysics-based model for approximating fitness effects of mutations. Though limited in its ability to describe all mutations, here we argue that our model provides the first simple, bottom-up approach to understanding mutational fitness effects.

We strengthened our protein model by merging it with stochastic population genetics simulations that include polyclonality, genetic drift, and linkage between sites. In our simulations, model proteins are continually buffeted by mutations that usually undermine protein stability. These destabilizing mutations shift the folded-unfolded equilibrium toward unfolded proteins, which imposes a context-dependent, usually small, fitness penalty proportional to the extent of unfolding. Partially destabilized proteins can be compensated by subsequent stabilizing mutations that replenish the fraction of folded protein and improve fitness. The asexual population dynamics yields a steady state distribution of protein stabilities ($p(\Delta G)$), from which we obtain a DFE with the same qualitative shape as observed experimentally. We find that $p(\Delta G)$ shifts toward instability for high mutation rates and small population sizes. This shift in stability disperses the DFE, i.e. increases the absolute selection coefficient of mutations and decreases robustness. Although the principles of our model are applicable to all species, we focus on viruses due to their relative simplicity and their extensively measured DFE (7-11).

## *Results*

**Nearly neutral thermodynamic fitness landscape.** Model genomes comprise $\Gamma$ genes, each encoding an essential protein completely described by a free energy of folding ($\Delta G_i$). We do not explicitly represent nucleotides. Many real proteins, or their domains, fold "2-state" (12) and fluctuate in thermal equilibrium between a unique native conformation and a set of unfolded "decoy" conformations. By elementary statistical mechanics, the fraction of time spent in the native state ($P^{nat}$) is given by

$$P^{nat} = \frac{1}{1+e^{\Delta G/k_bT}} \qquad [1],$$

where $k_b$ is Boltzman's constant and $T$ is temperature. We assume that native structures are perfectly functional while all unfolded conformations are completely nonfunctional. Thus, $P^{nat}$ is proportional to the concentration of functional protein. This paradigm assumes that function requires well-defined structure, which is not true for "intrinsically disordered" proteins (13).

While viral proteins often have short disordered regions, longer stretches, e.g. disordered domains, exist mostly in eukaryotes (14).

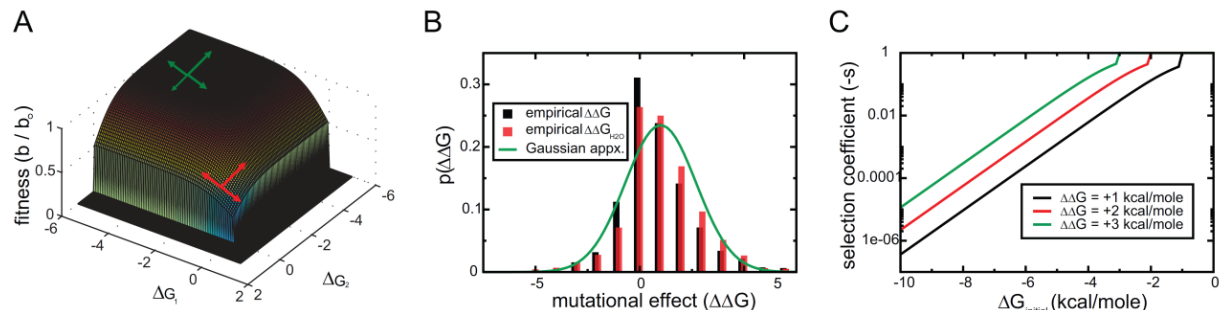

**Fig. 1** Nearly neutral thermodynamic fitness landscape. (A) A 2D section of the $\Gamma$ dimensional fitness landscape (based on Eq. 2). Highly stable genomes are robust to mutations (green arrows), whereas mutations to marginally stable proteins strongly impact fitness (red arrows). If $\Delta G>0$ for any gene, we set fitness to zero (lethal phenotype). (B) The distribution of thermodynamic mutational effects ($p(\Delta\Delta G)$). We approximate $p(\Delta\Delta G)$ by a Gaussian distribution, in rough agreement with ~4000 biophysical measurements from the ProTherm database (15). ≈28% of mutations are stabilizing and therefore technically beneficial, but most of these will have negligibly small fitness effects. C) The selection coefficient ($s \equiv \Delta b/b$) against representative destabilizing mutations increases exponentially as proteins lose stability (eq. 3). The kinks observed for $\Delta G_{initial}= -1, -2, -3$ occur because of lethal mutations that push $\Delta G>0$.

Viability requires that all essential proteins fold and have their native activity. This suggests the AND-like fitness function:

$$b = b_o \prod_{i=1}^{\Gamma} P_i^{nat} \qquad [2],$$

where $\Gamma$ is the number of essential genes and $b$ is the birth rate (i.e. fitness). $b_o$ summarizes the activity of all proteins and equals zero (lethal phenotype) if a mutation abolishes activity of any protein (see below). Truly beneficial mutations would correspond to increases in $b_o$, but we focus on short evolutionary timescales during which these mutations are unlikely to occur. Thus, without loss of generality, we henceforth scale time such that $b_o=1$ for all viable viruses. Eq. 2 states that fitness is reduced by the partial *absence of folded* protein. In reality, fitness is also compromised by the *presence of unfolded* proteins, which tend to aggregate and poison the organism. We heuristically capture this effect by further assuming that lethal phenotypes occur whenever $P_i^{nat} <0.5$ (i.e. $\Delta G_i > 0$) for any gene. The resulting fitness function is illustrated in fig. 1A. Note that we do not assume any tradeoff between fitness and stability. Such tradeoffs often result from mutations to an enzyme's active site (16-18) when evolving a new function, i.e. when *adapting*, which we do not consider.

We consider two types of nonsynonymous mutations. The first type abolishes activity by introducing STOP codons or disrupting critical residues (e.g. the active site), leading to $b_o=0$. We conservatively estimate that, together, these *unconditionally lethal* mutations comprise 10% of all nonsynonymous mutations (see supporting information (SI) *Text*). The remaining 90% alter thermodynamic stability (and thus $b$) by an amount $\Delta\Delta G$ drawn from a Gaussian distribution ($p(\Delta\Delta G)$) with mean +1 kcal/mole and standard deviation 1.7 kcal/mole (19). Our

Gaussian form closely approximates *p(ΔΔG)* obtained computationally (20), as well as thousands of biophysical measurements taken from the ProTherm database (15) (fig. 1B). Since *(ΔΔG)$_{mean}$ ≈ 1* kcal/mole, a protein with stability *ΔG* can tolerate *ΔG* mutations, on average, before unfolding. Mutations that push *ΔG>0* are considered lethal (see above). Note that relatively stable proteins have more viable single mutant "neighbors" than do less stable proteins (21). Additionally, we assume that *p(ΔΔG)* is (i) independent of *ΔG* and (ii) the same for all proteins. Both (i) and (ii) are roughly supported by computational studies (20, 22). Of course, (i) must fail for some proteins, e.g. the most stable sequence folding to a particular structure, but hyper-stable sequences are sufficiently rare that they are not visited by evolution in simulations or observed in reality (15).

We refer to eqs. 1,2, along with our Gaussian form of *p(ΔΔG)*, as the "nearly neutral thermodynamic landscape," which features continuous fitness effects (Fig. 1C) and complex epistatic patterns (figs. 1, 2). A natural measure of mutations' strength is the selection coefficient: *s≡ ($b_{after}$ - $b_{before}$)/$b_{before}$*. Note that *s* is independent of "bystander" proteins not involved with the mutation, since those factors of $P^{nat}$ cancel in eq. 2. Using eq. 1, it is easy to show (SI *Text*) that

$$s \sim e^{\Delta G/k_b T}(1-e^{\Delta \Delta G/k_b T}) \quad [3],$$

which is plotted in fig. 1C. Eq. 3 has two important consequences. First, since *ΔΔG* is Gaussian, *s* is log-normally distributed (for a single gene), featuring long tails of deleterious mutations (fig. S1) when *ΔG<<0*.

A second important consequence of eq. 3 is that the landscape is epistatic, since *s* depends not only on mutations' biophysical impact (*ΔΔG*), but also on protein stability (*ΔG*) prior to mutation. We will fully explore epistasis in future work—for now we merely point out important features. First, although destabilizing mutations are *energetically* independent, their *fitness* effects interact synergistically when they occur in the same gene. This "negative epistasis for fitness" follows from the sigmoidal shape of eq. 1, which itself follows from cooperative protein unfolding. Recently, negative epistasis was also observed in yeast regulatory sites using an energy based approach (23). Secondly, our fitness landscape appears smooth when defined over the space of protein stability (fig. 1A), but is actually *rugged* when viewed in sequence space because of the random, statistical way that we assign mutational effects via *p(ΔΔG)*. This ruggedness is evident in fig. 2, where colored curves depict random walks through sequence space that terminate upon lethal mutations. Compensatory mutations occur frequently. See fig. 2 caption for further commentary.

When applied to a single gene, our landscape is verified experimentally with *no adjustable parameters*. Refs. (24, 25) introduced mutations by in vitro mutagenesis of an antibiotic resistance enzyme (*TEM-1* β-lactamase), which was then expressed in *Escherichia coli*. Fitness was measured as the fraction of clones surviving antibiotic treatment, averaged over all clones carrying a common number of mutations. Using eqs. 1,2 and *Γ=1*, our model *quantitatively* matches the observed decline in enzyme function (fitness) as the number of mutations (*k*) increases (fig. 2). For small *k*, fitness declines exponentially due solely to unconditionally lethal mutations. For larger *k*, after thermodynamic stability is exhausted, the decline becomes steeper. An approach by Bloom et. al (24) that approximates protein function by a binary variable (0 or 1) also matches the averaged data. However, such landscapes (which we refer to as "strictly neutral") preclude both compensatory mutations and continuously varying fitness effects (fig. 2).

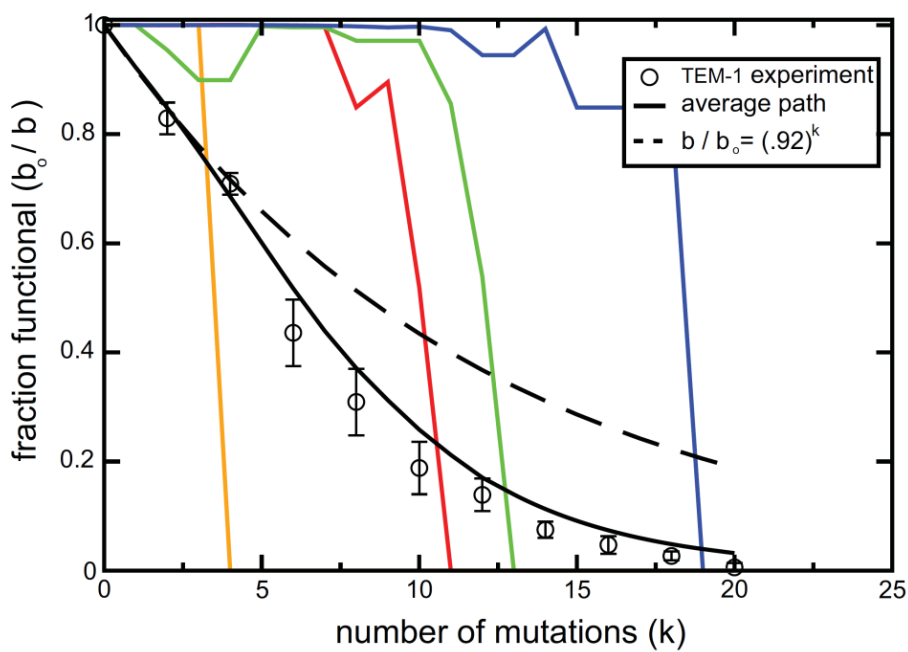

**Fig. 2** Unbiased mutation accumulation in a single gene, starting at $\Delta G=-8$ kcal/mole (26) (see SI *Text*). Fitness was evaluated using eq. 2, setting $\Gamma=1$. Colors: random walks through sequence space meander atop the plateau in fig. 1A until acquiring an unconditionally lethal mutation or eventually encountering the "cliff" at $\Delta G=0$. Compensatory mutations occur frequently. The generically concave-like paths reflect synergistic (i.e. negative) epistasis among deleterious mutations. By contrast, the average (solid black curve) of 10,000 random walks masks the synergistic epistasis and misrepresents the fitness interactions between particular mutations. Our approach matches experiments performed on *TEM-1* β-lactamase (circles). For comparison to this specific protein, we assumed that 8% of all mutations (including synonymous) are unconditionally lethal and that 30% are synonymous (25). The dotted line shows exponential decay with rate 8%. $k_bT = 0.62$ kcal/mole throughout. Otherwise, there are no free parameters. Data is from ref. (25) at 12.5 μg/mL ampicillin.

**Distribution of fitness effects (DFE).** The collection of first mutational steps in fig. 2 encodes the DFE for a particular gene: *TEM-1* β-lactamase. Recently, data from ref. (25) was reused to estimate the DFE for this enzyme (27). The resulting DFE was mostly unimodal, with ≈60% of mutations neutral, 8% lethal, and none beneficial/compensatory. While that result is valuable, the question remains as to how mutations impact the fitness of *arbitrary* proteins and the organism (virus) as a whole. To address this question in the context of our nearly neutral thermodynamic landscape, we must first possess the distribution of stabilities ($p(\Delta G)$) among all of the virus's (essential) proteins. One approach is to use the global, empirical ($p(\Delta G)$) from ProTherm (15) (see SI *Text*). However, in general, we predict that $p(\Delta G)$ depends on the evolutionary forces of mutation, selection, and stochastic drift (see below). To this end, we prepared the desired distribution $p(\Delta G)$ by simulating *populations* of stochastically evolving asexual viruses in our near-neutral model (see Methods) each containing $\Gamma=20$ essential proteins, under a range of population sizes ($N$) and realistic mutation rates ($m$).

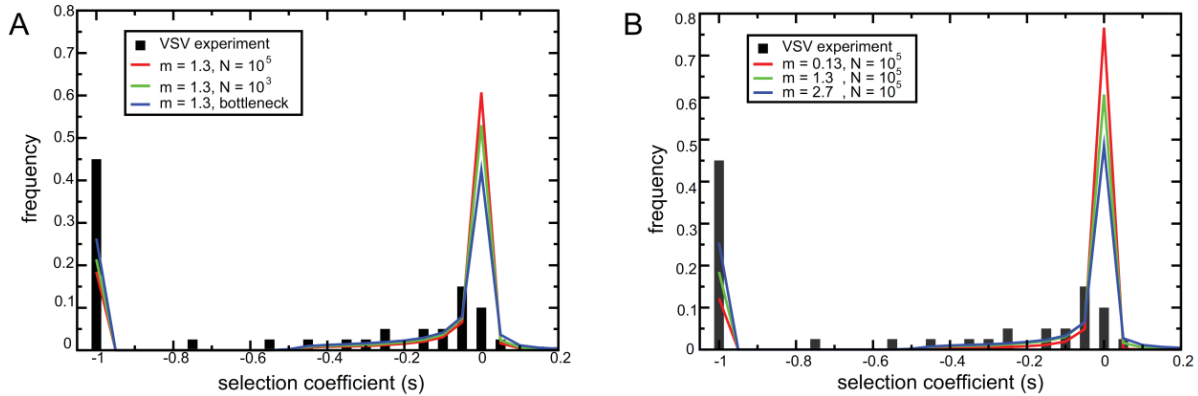

**Fig. 3** The distribution of fitness effects (DFE) due to random nonsynonymous mutations. Solid lines are from simulations, black bars are from experiments on VSV virus (7) (A) Dependence of the DFE on population size ($N$). Our model DFE becomes more dispersed as $N$ decreases or when the population experiences deep bottlenecks (see Methods). (B) Dependence of the DFE on mutation rates $m$. High $m$ also disperses the DFE, i.e. decreases mutational robustness. Bin widths are 0.05 for all data. We excluded five synonymous mutations from the original experimental dataset.

After evolving for a long time (50,000 generations), populations reached a steady state of mutation/selection/drift balance, independent of initial conditions (see SI *Movies*). Starting from steady-state, we separately mutated genes in each "virus" and measured the resulting distribution (DFE) of selection coefficients ($s$). We then averaged these distributions over the population and independent runs. Fig. 3 compares our DFE to an experiment on the ssRNA based vesicular stomatitis virus (VSV) in which 40 random nonsynonymous mutations were introduced by site-directed mutagenesis (7). Experiments on other viruses yielded qualitatively similar results (10) (fig. S2). The experimental and model DFE each feature a bi-modal shape, with a peak near neutrality ($s=0$) and another peak at lethality ($s=-1$). Interestingly, both model and experiment show some "beneficial" mutations ($s>0$), even though the population is not being challenged with a new environment. Our model interprets these as stabilizing mutations that *compensate* partially destabilized proteins rather than true, novel adaptations. Fig. 4 shows that our model can account for intermediate to large deleterious and compensatory fitness effects: e.g. non-lethal deleterious mutations decrease fitness by 2-12% on average. By contrast, on strictly neutral landscapes [e.g. refs. (19, 24, 28)], all mutations must be either lethal or neutral. Since the DFE for a single gene is log-normal, the overall model DFE is a linear combination of log-normals, with weights given by $p(\Delta G)$. This helps explain why the experimental DFE is easily fit by log-normal distributions (10, 11).

A striking feature of figs. 3,4 is that the DFE depends on population size ($N$) and mutation rate ($m$). In other words, the DFE depends not only on molecular constraints and the fitness landscape implied by them, but also on evolutionary forces that pull the population toward the flat region of fig. 1A or push it toward the "cliff." In particular, we find that mutations are more potent (i.e. genomes are less robust) in small populations with high mutation rates. We now analyze the origins of these effects by considering how a population's "location" on the fitness landscape, measured by $p(\Delta G)$, depends on $N$ and $m$.

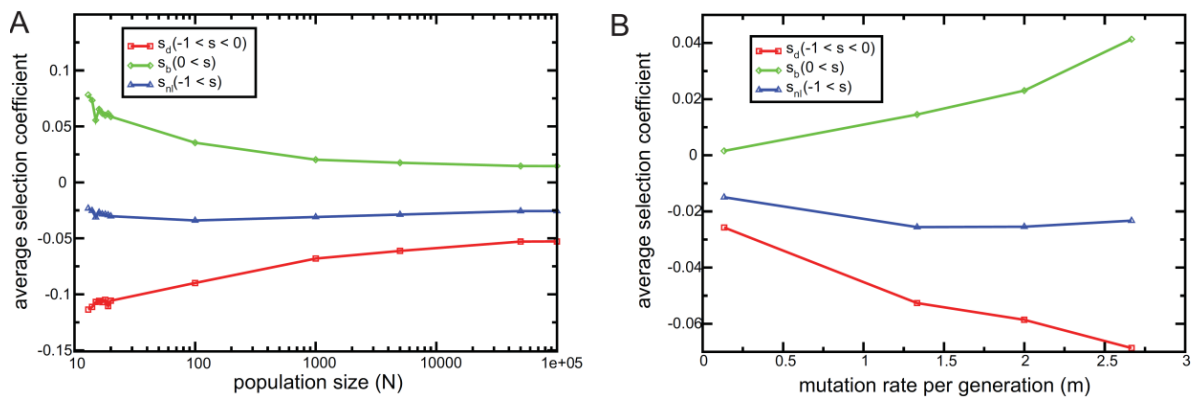

**Fig. 4** Dependence of selection coefficients on population size ($N$) and mutation rate ($m$). (A) The strength of mutations decreases with $N$ and (B) increases with $m$. Regardless of $N$ and $m$, mutations altering protein folding thermodynamics have substantial fitness effects, e.g. ≈2% among all non-lethal mutations. Parameters are $m=1.3$ (A) and $N=10^5$ (B).

**Small population size decreases evolved thermodynamic stability.** Fig. 5 illustrates the distribution of protein stability $p(\Delta G)$ for several values of $N$ and $m$. Each instance of $p(\Delta G)$ vanishes at $\Delta G=0$, peaks when $-\Delta G$ equals a few kcal/mole, and decays for more negative $\Delta G$. These qualitative features agree with previous work (analytic and simulation) (19, 28, 29) as well as the empirical distribution sampled from different genes and organisms in all kingdoms of life, obtained from the ProTherm database (15). For appropriately chosen $N$ and $m$, $p(\Delta G)$ from simulations agrees *quantitatively* with the global distribution from ProTherm (22) (fig. S3). We note that, since fitness increases monotonically (though exponentially weakly) with $-\Delta G$ in our model (eq. 1), perfectly fit proteins would have $\Delta G=-\infty$ which is obviously not observed. Therefore, the distribution's peak near $\Delta G \approx -2 - -5$ kcal/mole does *not* reflect optimality of this stability range, but rather a balance between mutation, selection, and random drift. Specifically, deleterious mutations with $|s| < 1/N$ have a significant probability of taking over the population (via random drift), making small populations especially poor optimizers (4). Fig. 5 shows that, in mutation/selection/drift steady-state, partially destabilized, i.e. less fit proteins are more heavily represented in small populations. Mutations that occur within these destabilized proteins have large fitness effects (fig. 1C) that disperse the DFE (figs. 3,4).

We can semi-quantitatively understand how $p(\Delta G)$ depends on $N$: In an ideal (mostly monoclonal) population, destabilizing mutations will cease to fix when $Ns \sim -Ne^{\Delta G/kT} \sim 1$, (4) i.e. when $\Delta G \sim -k_b T \ln N$. This expression matches our data reasonably well for $Nm < 1$ (fig. 5B). For larger $Nm$ (polyclonal populations), $N$ should be replaced by some "effective population size" $N_e$ (which is less than $N$), reflecting the additional stochasticity conferred by "background selection" (30). Evidently, populations as large as $N=10^5$ cannot be treated deterministically (i.e. $N \to \infty$), since $\Delta G$ continues to decrease with $N$ (fig. 5B). We also note that deep population bottlenecks (see Methods) abet protein destabilization in our simulations (fig. 5A), and therefore this effect is not specific to constant $N$ population genetic models.

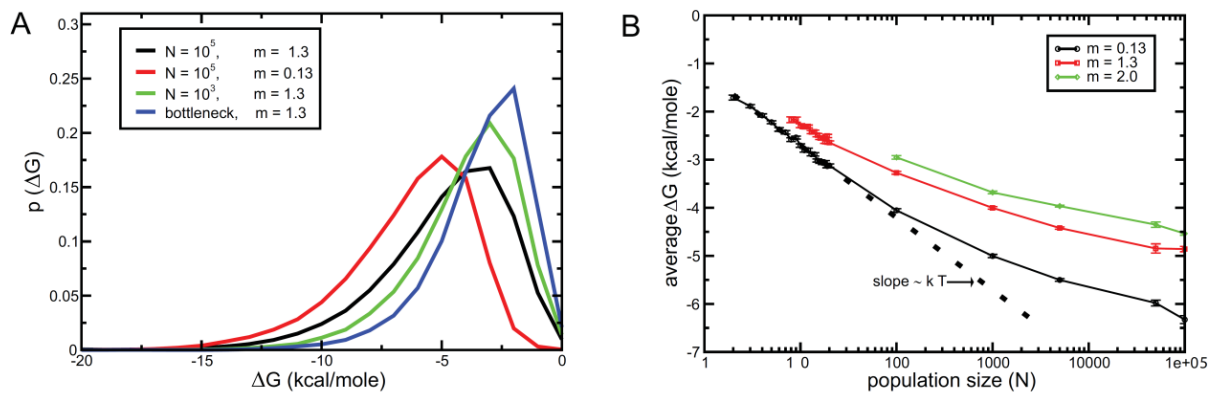

**Fig. 5** Protein stability increases with population size ($N$) and decreases with mutation rate ($m$). (A) The distribution of stabilities ($p(\Delta G)$) of all proteins in populations with different $N$ and $m$. Genetic drift and mutational load shift the distribution of evolved stabilities toward $\Delta G=0$. Deep population bottlenecks also decrease protein stability. (B) Dependence of $\Delta G$ (averaged over the population and >10 replicates) on $N$. For small $Nm$, the population is mostly monoclonal and $\Delta G$ decreases as $\Delta G \sim k_b T \ln N$. The dotted line is fit from the interval $2 \leq N \leq 20$ (slope = 0.65). Values of all parameters are stated on the figures/axes. Error bars are 1 s.e.m.

**High mutation rates decrease evolved thermodynamic stability.** Deleterious mutations, most of which are not completely lethal, are incessantly produced with rate $\sim m$. Meanwhile, prior to being purged from the population, deleterious mutations linger for $\sim b_{avg}/(b_{avg} - b) \sim 1/s$ generations, independent of $m$. Therefore, as $m$ increases, more deleterious mutations, in the form of slightly destabilized proteins, accumulate in the population. This reduction in fitness, known as the "mutation load," (31) is evident in fig. 5A and can be quite strong (e.g. $\approx 3.5$ kcal/mole for $N=10^5$, see fig. S5). Mutations that arise in the background of excess destabilized proteins have large fitness effects, thereby dispersing the DFE (figs. 3,4) and decreasing the robustness of genomes to mutations.

Although this phenomenon is grounded in simple population genetics principles, it clashes with several existing models of "neutral networks," (19, 21, 32, 33) in which a mutationally connected set of genomes is perfectly fit while another set is completely lethal. Those studies predict that at large $m$, sequences connected to many viable mutational neighbors will predominate over less connected sequences. In the language of our model, the prediction from neutral network theory is that greater stability will occur at large $m$, which contradicts fig. 5A. In supporting information (SI *Text*), we show that our model reproduces the neutral network prediction if we artificially set temperature ($T$) to zero, i.e. if we eliminate *all* curvature from our fitness landscape. This point is very important in understanding the general relationship between mutation rate and mutational robustness (34, 35).

## *Discussion*

Our nearly neutral landscape is based on molecular biophysics: elementary thermodynamics and the distribution of mutational effects ($p(\Delta\Delta G)$) on folding free energy (Fig.1B). Thus, evolutionary dynamics in our model are shaped by biophysics. Reciprocally, the physical properties of evolved proteins are shaped by evolutionary dynamics. In particular we predict that organisms with high mutation rates ($m$) and small (effective) population sizes ($N$) will evolve relatively unstable proteins (fig. 5) that reside on the landscape's steep gradient (fig. 1A). Thus, for these proteins, stabilizing mutations can have large compensatory effects while destabilizing mutations are often strongly deleterious or lethal (if $\Delta G>0$ after the mutation). Therefore, the distribution of fitness effects (DFE) becomes more dispersed for large $m$ and/or small $N$ (figs. 3,4).

Our most dramatic prediction is that the thermodynamic impacts of mutations can be inferred from their observed fitness effects. In recent site-directed mutagenesis experiments (10), the fitness effect and exact nucleotide change were recorded for each mutation. Therefore, one can purify the wild-type and mutant proteins, biophysically measure $\Delta G$ for each, and then *quantitatively* compare their relationship to that predicted by our model (eqs. 1,2).

Short of this direct confirmation, our approach has indirect experimental support. First, applied to a single gene, our model agrees quantitatively with in vitro mutagenesis experiments (24, 25) (fig. 2). The agreement with experiment is especially remarkable, considering that it involves no adjustable parameters (i.e. no curve fitting).

A second form of experimental support is that the distribution of protein stabilities ($p(\Delta G)$) from simulations closely approximates the global, empirical distribution from the ProTherm database (15, 19, 22) (fig. S3). This implies that we could have obtained the model DFE by using only data in ProTherm and avoiding evolutionary simulations altogether, assuming that the distribution of stabilities in each virus follows the ProTherm histogram (fig. S4). However, a major shortcoming of that route is that it cannot reveal the relationship between the DFE, $N$, and $m$ predicted by our model.

Thirdly, our predicted DFE has the same qualitative features observed experimentally for five viruses (7, 11) (fig. 3). Although our model overestimates the fraction of neutral mutations, the shape *and the scale* of mutational effects are qualitatively correct.

Apart from reproducing existing experimental observations, our model makes concrete, testable *predictions*. First, we predict that small populations have less stable/robust proteins. Experiments on bacteria show that decreased fitness caused by single cell bottlenecks (small effective $N$) naturally induces, and is partially compensated by, chaperone overexpression (36, 37). Thus, protein unfolding is implicated as a source of the fitness decline in those experiments. Additionally, proteins from the intracellular parasite (small $N$) *Buchnera aphidicola* were predicted computationally to be less stable than their homologs in free living relatives (38). While "strictly neutral" network theory also predicts this result (28), the underlying mechanism is different. Our landscape features nearly neutral mutations that escape selection, depending on $Ns$. By contrast, in the strictly neutral framework there is no "$s$," and everything depends only on the product $Nm$ (28, 32).

A second concrete prediction is that viruses with high mutation rates (RNA viruses) have less stable proteins than those with low mutation rates (DNA viruses). This prediction can be directly tested by biophysically measuring $\Delta G$ both in RNA viral proteins and their homologs in DNA viruses. Although this has not yet been done, a recent study (14) found a relatively low density of van der Waals contacts among RNA viral proteins, which suggests, but does not prove, that these proteins are less stable. A closely related prediction is that RNA viruses are less mutationally robust than DNA viruses. We predict that mutational robustness, in particular the fraction of non-lethal mutations, is diminished for species with high mutation rates. Accordingly, experiments often find more lethal mutations in RNA viruses (high $m$) than DNA viruses (10) (fig. S2).

Our prediction that robustness decreases with $m$ is at variance with previous theoretical work, e.g. neutral networks (21, 32, 33) (discussed previously and in SI *Text*). Aside from neutral networks, robustness has also been investigated in "digital organisms" which compete for CPU resources during adaptation on a complex fitness landscape (39). The main result from those studies is that digital strains evolved under high $m$ defeat those evolved under low $m$ when competition between the two strains occurs at high $m$. In this sense, high $m$ promotes robustness to mutations in digital organisms. By contrast, in our model, the low $m$ strain would always be victorious. The discrepancy in competitive outcome occurs because the digital high $m$ and low $m$ populations settled on completely different fitness peaks *during the course of adaptation*. By contrast, our model does not describe distinct fitness peaks or dramatic adaptation. However, insofar as *local* fitness peaks are concerned, our results agree completely with those from digital organisms, where the strength of single mutational effects increased with $m$, as in our fig. 4B (see fig. 2 from ref. (39)).

The relationship between robustness and $m$ has also been investigated experimentally. For example, inspired by the results from digital organisms (39), Sanjuán et. al showed that a more robust viral strain out-competed a less robust strain only in the presence of chemical mutagens (large $m$) (40). Some aspects of this finding are consistent with our predictions: Excess robustness, should it exist, is beneficial in our model, particularly when $m$ is large. However, we predict that excess robustness, if due to increased protein stability, is short-lived (41) (see SI *Movies*), and will not evolve in response to large $m$. The reason that excess robustness cannot be maintained, despite its advantage at high $m$, is that high $m$ also increases the "mutational wind," which decreases robustness. It remains to be seen whether depletion of excess robustness would occur experimentally with these two viral strains. Importantly, these strains had very complex evolutionary histories: one was a chimera of natural isolates whereas the other had recently adapted to a new host in the laboratory. Our model cannot take into account those complications or, consequently, explain all aspects of that experiment, e.g. why the less robust strain prevailed at low $m$.

In the future, our model can be generalized and improved in several ways. First, eq. 2 could take a more complicated, nonlinear form, thereby sharpening the landscape's gradient near $\Delta G=0$. Secondly, we assumed that each gene has the same expression level. In reality, highly expressed genes may cause lethality when only mildly destabilized, because aggregation and toxicity depend on the *number*, not the fraction, of unfolded molecules. Thirdly, for simplicity, we have assumed that a fixed fraction (10%) of nonsynonymous mutations are unconditionally lethal. Careful inspection of real viral proteins could improve this estimate. Fourthly, we implicitly assumed a low multiplicity of infection (MOI) of host cells. Real viruses often have high MOI, in which one virus's phenotypic defects are complimented by another virus in the same cell (42). This effect can relax selection and possibly reduce protein stability. Fifthly, chaperones expressed by hosts, as well as structurally disordered regions, could complicate our simple description of protein folding. In particular chaperones can mitigate deleterious fitness effects of destabilizing mutations up to a certain degree, thus sharpening the "cliff" in the fitness landscape. Finally, our model does not explicitly include protein-protein interactions, the basic features of which could be included in future work (43). Any or all of these limitations might improve the quantitative agreement between the model and experimental DFE. Of course, the most obvious limitation is that real proteins can be disrupted in ways that our model cannot predict. Nevertheless, we have demonstrated that insight to the origin of fitness effects can be gained by our model, which combines biophysical realism with the computational and conceptual minimalism of a "toy model."

## *Methods*

In simulations, a virus is chosen randomly with weight $b$ to replicate in continuous time. Each replication event represents an "infection cycle," which in reality often entails several genome replications. Assuming random codon usage, we set 1/4 of mutations as synonymous (44), which are ignored by the simulations. Upon replication, both parent and daughter genomes (semi-conservative replication) independently acquire a random number of mutations drawn from a Poisson distribution with mean $m$. If the number of viruses exceeds a fixed threshold ($N$), a randomly selected individual is removed. Individuals carrying lethal mutations are immediately removed. Since replication is semi-conservative, both parent and offspring can acquire a lethal mutation, thereby possibly reducing the size below $N$. For sufficiently high mutation rates, the population may in fact go extinct (22). All simulations in this work were done with sufficiently small $m$ so that the population size almost always equaled $N$. We expect that unconditionally lethal mutations minimally affect $p(\Delta G)$, though they do impact the DFE. To minimize equilibration times, all simulations were initialized with a clone whose genome was drawn from the analytical estimate of the steady-state stability distribution from ref. (19). One "generation" denotes $N$ infection cycles (i.e. birth events). Note that this scheme is similar to the standard Moran process (31). Bottlenecked populations evolved for $\approx$8,700 repeated cycles ($10^5$ generations) of growth to $N=10^5$ followed by sampling down to $N_{min}=10$.

## *Supporting Information*

### Selection coefficient increases exponentially with ΔG and is approximately log-normally distributed for a single gene

$$s \equiv \frac{b_{after} - b_{before}}{b_{before}} = \frac{P_{after}^{nat} - P_{before}^{nat}}{P_{before}^{nat}} = \frac{P_{after}^{nat}}{P_{before}^{nat}} - 1 = \frac{1 + e^{\Delta G_{before}/kT}}{1 + e^{\Delta G_{after}/kT}} - 1$$

If $exp(\Delta G_{after}/k_b T) \ll 1$, i.e. $\Delta G_{after} \ll 0$, then

$$s \approx (1 + e^{\Delta G_{before}/kT})(1 - e^{\Delta G_{after}/kT}) - 1 = e^{\Delta G_{before}/kT}(1 - e^{\Delta\Delta G/kT} - e^{\Delta G_{before}/kT}) \approx e^{\Delta G_{before}/kT}(1 - e^{\Delta\Delta G/kT})$$

Therefore,

$\Delta\Delta G = 1 - se^{-\Delta G_{before}/kT}$ (i.e. a linear transformation of *s*) is log-normally distributed since $\Delta\Delta G$ is Gaussian. Fig. S1 compares this approximation to the DFE obtained for single genes via simulation. This approximation neglects lethal mutations (both unconditional ones and those pushing $\Delta G > 0$). A log-normal distribution has been used previously to fit the *overall* experimental DFE (10, 11).

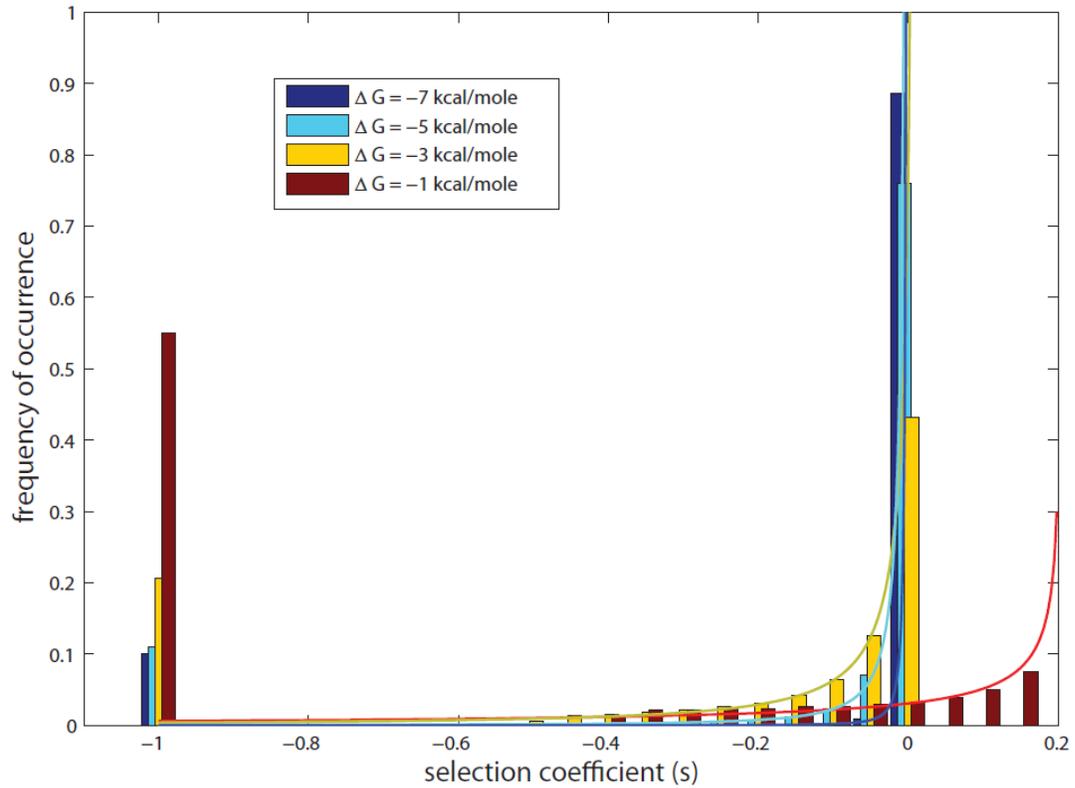

Fig. S1. The DFE for single genes is approximately log-normally distributed. Clearly, less stable genes have far more dispersed DFE. The "exact" data (bars) resulted from $10^6$ random mutations to genes of the indicated stability. Although the analytic estimates (lines) perform reasonably well, they do not account for lethal mutations, resulting in some error in the deleterious tail.

**Experimental DFE of nonsynonymous mutations for five viral species.** We compiled experimental results from refs. (7-9, 11) and removed non-synonymous and intergenic mutations (which our model clearly do not capture). The results are shown in Fig. S2.

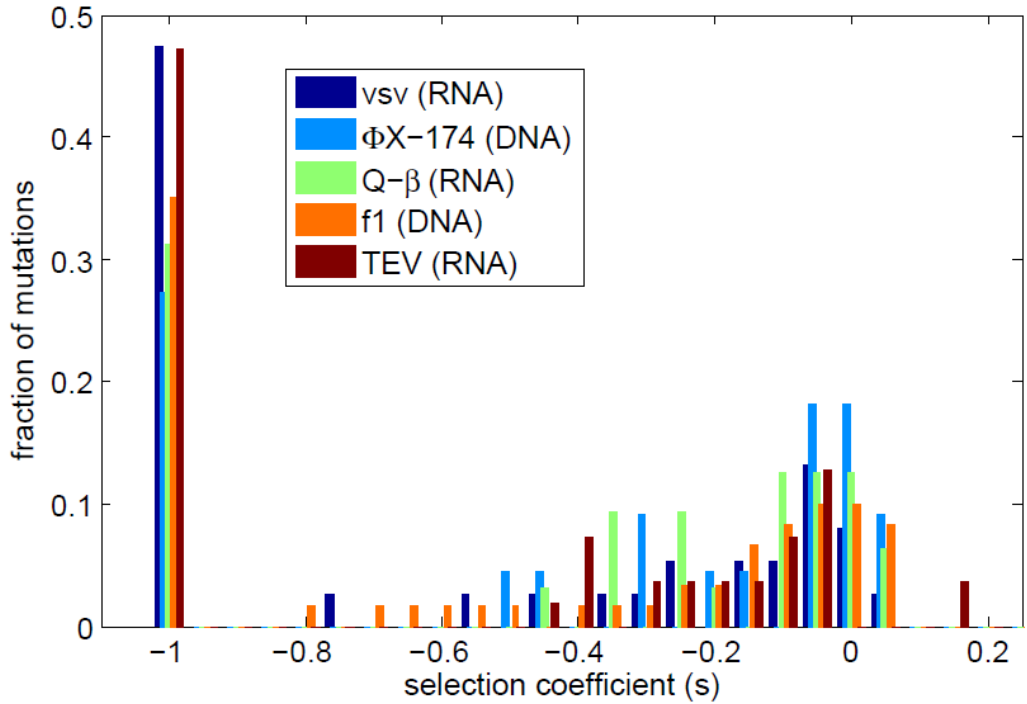

Fig. S2. Experimental DFE for 5 viruses, obtained by site-directed mutagenesis (7-9, 11). We removed all synonymous and intergenic mutations from the raw data. The fitness measure for the TEV data (8) were transformed according to the formula provided by Sanjuán (10). The DFE shape is similar across the five species, and RNA viruses tend to have more lethal mutations.

**Empirical distribution of protein stabilities.** The ProTherm database (15) contains 421 stability measurements from different proteins and phylogenetically diverse species. Clearly, we cannot assign a single population size (N) and mutation rate (m) to this dataset. Nevertheless, fig. S3 shows that the distribution of experimental $\Delta G$ measurements ($p(\Delta G)$) agrees with results from our simulations. Furthermore, fig. S4 shows that the DFE computed using the empirically obtained $p(\Delta G)$ is similar to that obtained via simulation (main text). Thus, evolutionary simulations are not required to deduce the basic form of the DFE. However, those simulations are required to reveal how the DFE depends on N and m (main text).

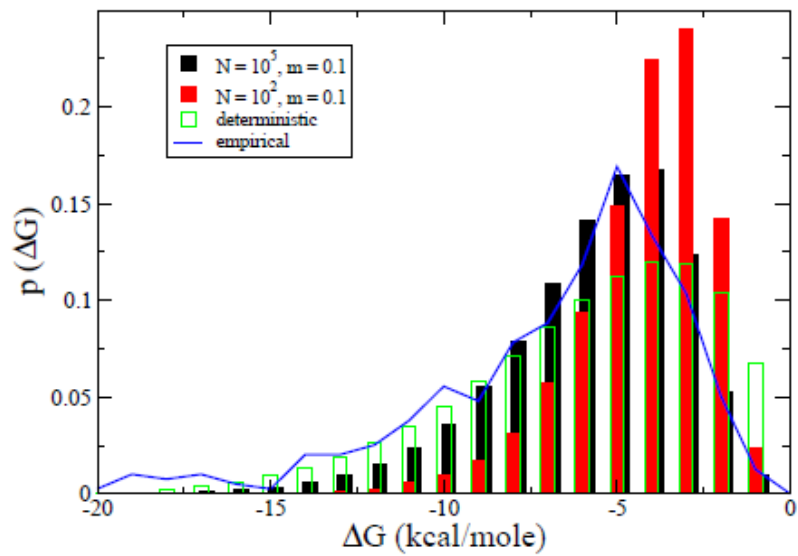

Fig. S3. P(ΔG) from simulations matches experimental values taken from the ProTherm database (15). This agreement was previously reported (19, 22). The analytic result derived in ref. (19), using a strictly neutral landscape, is also shown.

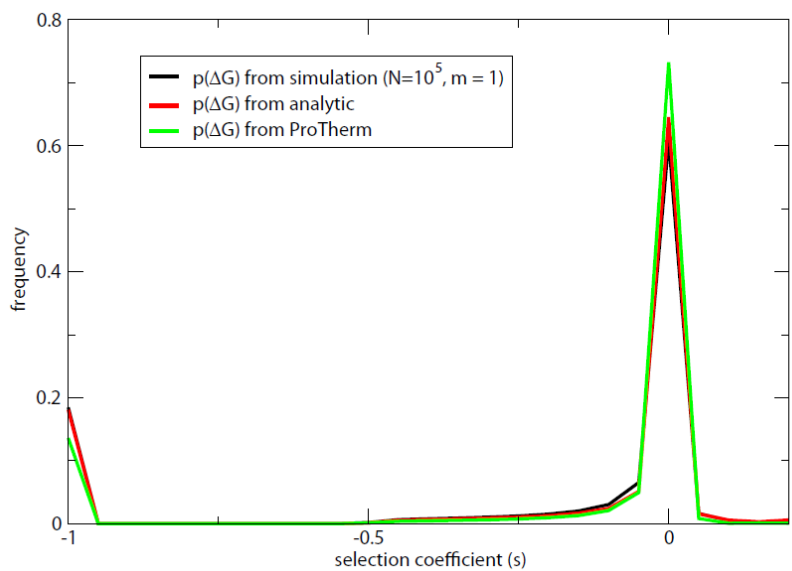

Fig. S4. The general properties of the DFE can be deduced from analytic approximations (19), from the ProTherm database (15), or from simulations, (as in the main text). In each case, we used our Gaussian approximation of p(ΔΔG), along with p(ΔG) from either ref. (19), Protherm, or the simulations in the main text. Of these three routes, only the one we employed in the main text can reveal how the DFE depends on N and m.

**Unconditionally lethal mutations.** The fraction of mutations that introduce STOP codons can be estimated from the genetic code. There are 61 × 9= 549 possible mutations from the 61 sense codons. Of these, 23 lead to stop codons (44), corresponding to ≈ 4.2% all mutations.

Mutations in an enzyme's active catalytic site also typically disable function. Assuming 3 catalytic residues, 3 nearby critical residues, and a 100 amino acid domain, 6% of all nonsynonymous mutations could plausibly disable the active site.

Thus, 6%+4.2% ≈ 10% is a reasonable conservative estimate for unconditionally lethal mutations. Note that this estimate omits insertions and deletions. Including these mutations would increase the fraction of lethal mutations and are DFE and improve agreement with experiment.

**Thermodynamic stability of wild-type TEM-1 β-lactamase.** This protein actually folds "3-state," with ΔG between native and intermediate states equal to 7.27 kcal/mole and ΔG between intermediate and unfolded states equal to 3.78 kcal/mole(26). These were each measured at T=25°C. Taking the effective two-state ΔG as the sum of these two values gives $\Delta G_{25}$=11.0 kcal/mole. However, the experiments relevant to fig. 2 from the main text were performed at 37°C (25). Since ΔG=ΔH − TΔS (where ΔH is the enthalpy and ΔS the entropy of folding/unfolding), the difference in temperatures shifts ΔG by an amount ΔT ΔS. Approximating ΔS ≈ 0.25/(mole °C) (45), we obtain $\Delta G_{37}$≈-11 kcal/mole + 3 kcal/mole = -8 kcal/mole, as in the main text.

**Discrepancy between nearly neutral and strictly neutral network predictions.** Our model predicts that mutational robustness decreases with *m*, whereas (strictly) neutral network theory predicts the opposite. Here, we pinpoint the source of this discrepancy. Our fitness landscape can be interpreted as a "nearly neutral network" that becomes "strictly neutral" in the hypothetical limit of zero temperature (*T=0*). In that case, eq. 1 becomes a Heaviside step function and all proteins with *ΔG<0* form a neutral network. When *T=0*, proteins with *ΔG≈0* experience no selective disadvantage, which pushes *p(ΔG)* far to the "right," relative to the *T>0* case (fig. S5A). Once pressed against the wall at *ΔG≈0*, *T=0* populations generate many lethal mutations, but these are *instantaneously* removed from the population and thus do not show up in *p(ΔG)*. Meanwhile, the minority of stabilizing mutations *do* appear in *p(ΔG)*. Thus, increasing *m* on the *T=0* landscape can only increase average protein stability. Indeed, in agreement with neutral network studies (21, 32, 33), we found that increasing *m* improved stability to a small extent when *T=0* but undermined stability to a larger extent when *T>0* (fig. S5B).

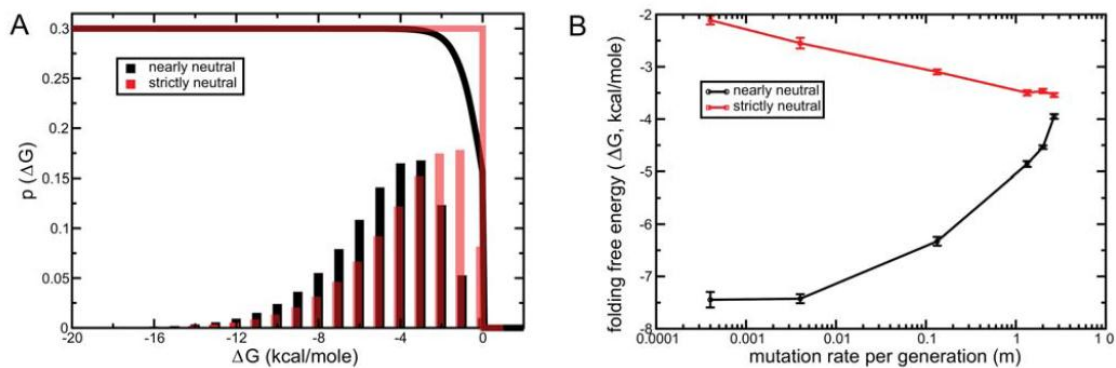

**Fig. S5** Comparison of nearly neutral and strictly neutral landscapes. Distributions of protein stabilities ($p(\Delta G)$) in populations evolved on strictly neutral (red) and nearly neutral (black) landscapes. Parameters are $m=1$ and $N=10^5$. Curves drawn above the histograms depict 1D sections of the landscapes upon which the populations evolved. Dependence of $\Delta G$ (averaged over populations and > 10 replicates) on $m$. Nearly neutral and strictly neutral landscapes yield opposite trends. Populations often went extinct for higher mutation rates. $N=10^5$. Error bars are 1 s.e.m.

## *Acknowledgements*

We thank G. Frenkel, N. Delaney, D. Rosenbloom, D. Lorenz, and especially A. Serohijos and S. Bershtein for comments on the manuscript. This work was supported by the National Institute of Health.